# Complementary lateral-spin-orbit building blocks for programmable logic and in-memory computing

*Nan Zhang[†], Yi Cao[†], Yucai Li, Andrew W. Rushforth, Yang Ji, Houzhi Zheng, and Kaiyou Wang\**

N. Zhang, Dr. Y. Cao, Y. Li, Prof. Y. Ji, Prof. H. Zheng, Prof. K. Wang
State Key Laboratory for Superlattices and Microstructures
Institute of Semiconductors
Chinese Academy of Sciences
Beijing 100083, China
E-mail: kywang@semi.ac.cn

Dr. Y. Cao, Prof. K. Wang
Beijing Academy of Quantum Information Sciences
Beijing 100193, China

Dr. A. W. Rushforth
School of Physics and Astronomy
University of Nottingham
Nottingham NG7 2RD, UK

N. Zhang, Y. Li, Prof. Y. Ji, Prof. H. Zheng, Prof. K. Wang
Center of Materials Science and Optoelectronic Engineering
University of Chinese Academy of Science
Beijing 100049, China

Prof. K. Wang
Center for Excellence in Topological Quantum Computation
University of Chinese Academy of Science
Beijing 100049, China

[†]These authors contributed equally to this work;
[*]Corresponding e-mail: kywang@semi.ac.cn

Keywords: spin-orbit logic, lateral spin-orbit torque (LSOT), current-driven magnetization switching, stateful logic, in-memory computing.

Current-driven switching of nonvolatile spintronic materials and devices based on spin-orbit torques offer fast data processing speed, low power consumption, and unlimited endurance for future information processing applications. Analogous to conventional CMOS technology, it is important to develop a pair of complementary spin-orbit devices with differentiated magnetization switching senses as elementary building blocks for realizing sophisticated logic functionalities. Various attempts using external magnetic field or complicated stack/circuit designs have been proposed, however, plainer and more feasible approaches are still strongly desired. Here we show that a pair of two locally laser annealed perpendicular Pt/Co/Pt devices with opposite laser track configurations and thereby inverse field-free lateral spin-orbit torques (LSOTs) induced switching senses can be adopted as such complementary spin-orbit building blocks. By electrically programming the initial magnetization states (spin down/up) of each sample, four Boolean logic gates of AND, OR, NAND and NOR, as well as a spin-orbit half adder containing an XOR gate, were obtained. Moreover, various initialization-free, working current intensity-programmable stateful logic operations, including the material implication (IMP) gate, were also demonstrated by regarding the magnetization state as a logic input. Our complementary LSOT building blocks provide an applicable way towards future efficient spin logics and in-memory computing architectures.

## 1. Introduction

For more than half a century, conventional microelectronic logic circuits based on complementary metal-oxide-semiconductors (CMOS), i.e. the electron (n-) and the hole (p-) type charge conduction devices, have been developed to assemble the present von-Neumann computing architecture. Generally, the information represented by charge carriers are volatile, which has to be transported frequently between the logic processing unit and the memory devices, and thereby consuming massive unnecessary powers while generating undesirable joule heating. As a promising solution to these problems, spintronic devices that utilize the nonvolatile electron spins in a ferromagnet have been suggested by the community over the past decades[1-3]. Particularly, technologies of spin-transfer torque (STT)[4] and then spin-orbit torques (SOTs)[5-7] not only offer fast data processing speed and low power consumption, but also provide capabilities of programmable spin-logic operations[8-9] as well as non-von-Neumann in-memory computing applications[10-11]. Analogous to CMOS technology, it is important to develop complementary spintronic logic building blocks[12-15], i.e. two type of basic spintronic devices that response distinctly to the same input signal, for facilitating complex logic functions with simplified circuit design.

Typically, the SOT-induced magnetization switching with perpendicular magnetic anisotropy requires the assistance of an in-plane external magnetic field[5], the direction and the magnitude of which determine the switching direction and the critical switching current density. Inspired by this unique feature of SOT switching, naturally, approaches of external magnetic field-dependent complementary spin-orbit logic devices have been proposed[8,16]. Recently, more scalable SOT technologies with external magnetic field-free switching have also been successfully demonstrated, and the magnetization switching direction can be controlled by various methods, such as introducing a build-in in-plane exchange magnetic field[17-18] and adjusting its direction, creating a spin current gradient[19-20] and tuning its polarity, manufacturing a lateral wedge oxide[21-22] and engineering its tilting orientation, and so on. Following these ways, field-free complementary spin-orbit logic pairs can be reasonably proposed, however, problem of either the existence of an unscalable in-plane coupling FM layer, or the fussy multi-terminal (terminal number > 3) SOT-MTJ device design, or the incompatibility with standard magnetic tunnel junction (MTJ) as well as the difficulties in manufacturing procedure, makes those potential complementary spin-orbit logic proposals not applicable for industrial realization. Thus, magnetic field-free complementary spin-orbit logic pairs with integration-friendly approaches are strongly desired.

Recently, a novel lateral spin-orbit torques (LSOT) induced field-free deterministic magnetization switching has been demonstrated in a locally laser annealed perpendicular magnetic anisotropy (PMA) Pt/Co/Pt structure, the switching orientation is dependent on the relative local annealing location of the in-plane current (for example, along $x$ direction) and the laser track (also along the $x$ direction, but lies on either -$y$ or +$y$ side of the sample)[23]. Inspired by this integration-friendly approach, here we show how a pair of magnetic field-free complementary LSOT logic devices can be demonstrated as building blocks for programmable and stateful logic operations. By setting the polarity of initialization electric current, basic Boolean logic gates of AND/OR (NAND/NOR) were programmed in a single -$y$ (+$y$) side laser annealed LSOT device, and an half adder containing the nonlinear separated XOR logic gate was realized by the combination of three such devices. Moreover, when regarding the magnetization state as a logic input, various initialization-free stateful logic gates were performed and programmed by adjusting the working current intensity. The demonstrated versatile logic functionalities based on our complementary LSOT building blocks provide an integration-friendly way to engineer future efficient spin logics and in-memory computing architectures.

## 2. Initialization current-programmable Boolean logics using complementary LSOT devices

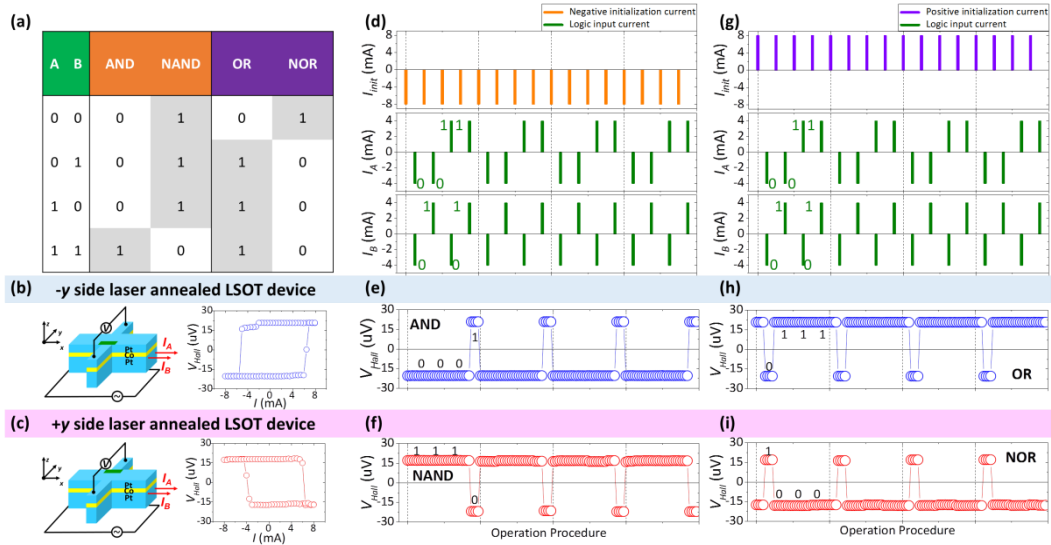

**Figure 1 | Initialization current-programmable Boolean logics operating on the complementary LSOT devices. a,** Truth tables of AND, NAND, OR, and NOR Boolean logic gates. **b-c,** Schematic drawings and respective *x*-direction channel pulse current-induced $V_{Hall}$-$I$ loops of the complementary LSOT devices. Black arrows indicate the Cartesian (*x, y, z*) coordinate systems. Both devices are Hall bars with an identical stack structure of Si/SiO$_2$ substrate/Pt(3 nm)/Co(0.5 nm)/Pt(2 nm), however, with different locations of laser annealing tracks (green zones) at (**b**) −*y* side and (**c**) +*y* side of the Hall crosses, respectively. The Hall bar channel width is 4 μm. The duration of every current pulse is 10 ms and the anomalous Hall voltage $V_{Hall}$ was obtained under a small d.c. current (100 μA) 1 second after each pulse. **d-i,** Demonstration of initialization current-programmable Boolean logic gates using the above two devices. Binary logic inputs of two current pulses $I_A$ and $I_B$ (-/+4 mA stands for logic value '0'/'1') were applied along the Hall bar channel simultaneously.

The resulting nonvolatile current-induced magnetization down/up state represented by the negative/positive $V_{Hall}$ was regarded as logic output value '0'/'1'. A necessary initialization current $I_{init}$ of -8 mA (orange pulses in (**d**)) or +8 mA (purple pulses in (**g**)) was applied on the devices before each operation, the polarity of which defined the type of the logic gate. For $I_{init}$ = -8 mA, the −$y$ and the +$y$ side locally laser annealed devices showed (**e**) AND and (**f**) NAND gates, respectively. Meanwhile, for $I_{init}$ = +8 mA, the −$y$ side and the +$y$ side locally laser annealed devices showed (**h**) OR and (**i**) NOR gates, respectively. The $x$-axis of (**d-i**) are operation procedures with same scales and values.

Figure **1a** shows the truth table of four common Boolean logic gates AND, NAND, OR, and NOR, where AND and NAND (also OR and NOR) are complementary function pairs that always output contrary results for same inputs. To facilitate these complementary functions, complementary devices with opposite switching orientations (anticlockwise/clockwise $V_{Hall}$-$I$ loops) are demonstrated. As shown in Figure **1b** and **1c**, two series of PMA Si/SiO$_2$ substrate/Pt (3 nm)/Co (0.5 nm)/Pt (2 nm) Hall bar samples with different locally laser annealing configurations (laser tracks on the -$y$ and the +$y$ side of the Hall cross, respectively) were fabricated. In the absence of an external magnetic field, deterministic LSOT induced current-driven anticlockwise/clockwise magnetization switching was observed in the -$y$/+$y$ side locally laser annealed devices. This dependence of the field-free switching orientation on the laser track location can be attributed to the polarity of LSOTs arising from the lateral Pt-Co asymmetry after laser annealing[23].

However, single-device implementations of $n$ different logic gates require devices with at least totally $n$ possible different modes. Hence, in order to realize the 4 logic gates shown in Figure **1a**, another binary variable, i.e. the initial magnetization state (spin down/up), is introduced to program the logic functions in each device, resulting 4 different single-device modes based on the complementary LSOT devices. Here, an initialization current pulse of $I_{init}$ = -/+ 8mA (upper panels of Figure **1d** and **1g**) was sufficient to reset the magnetization state before each logic operation. After the initialization, as shown in the middle and the lower panels of Figure **1d** and **1g**, two logic inputs represented by current pulses $I_A$ and $I_B$ with equal absolute values of 4 mA were applied along $x$ direction simultaneously, a negative/positive sign of which stood for the logic value '0'/'1'. In this way, three possible overlapped current pulses, i.e. $I_{ovlp}$ = -8 mA, 0 mA, and +8 mA were actually applied on the complementary LSOT devices with different initial magnetization states. As a result, various resulting logic outputs with nonvolatile magnetization states represented by $V_{Hall}$ under a 100 μA measuring d.c. current (spin down, negative $V_{Hall}$ for logic output '0'; spin up, positive $V_{Hall}$ for logic output '1') were shown in Figure **1d**, **1e**, **1h**, and **1i**. Remarkably, for initialization current $I_{init}$ = -/+8 mA, complementary Boolean logic gates of AND/OR and NAND/NOR were realized in the -$y$ and the +$y$ laser annealed LSOT devices, respectively. It is worth noting that the demonstrated NOR gate can also act as a NOT gate of input $I_A$ when $I_B$ is fixed at '0'.

## 3. Demonstration of a half adder by combining the LSOT devices

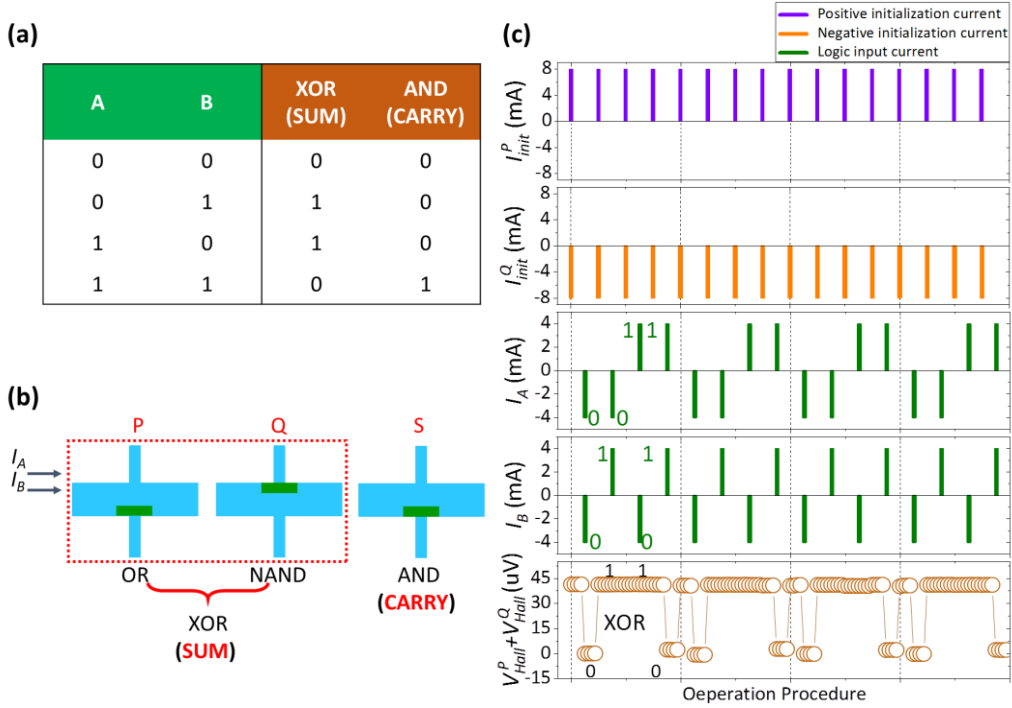

**Figure 2 | Demonstration of a spin-orbit half adder using three LSOT devices. a,** Truth table of a half adder, which can be divided into an XOR gate (working as SUM) and an AND gate (working as CARRY). **b,** Schematic drawing of the demonstrated half adder consisting of three LSOT devices denoted as P, Q, and S, where P and S are -*y* side laser annealed devices and Q is a +*y* side laser annealed device. By setting the initialization currents as $I_{init}^P$ = +8 mA, $I_{init}^Q$ =-8 mA, and $I_{init}^S$ =-8 mA, P, Q, and S are expected to function as OR, NAND, and AND gates, respectively. **c,** Realization of XOR by adding the logic outputs of OR P and NAND Q together, with regarding the almost cancelled near zero ($V_{Hall}^P$ + $V_{Hall}^Q$) as the XOR output '0' while the nearly doubled positive ($V_{Hall}^P$ + $V_{Hall}^Q$) as the XOR output '1'.

With these initialization current-programmable single-device Boolean logic gates,

more complicated spin logic functions can be substantially facilitated by optimized combinations of the complementary LSOT devices while programming their initialization currents. A practical case is the demonstration of a LSOT half adder, one of the core modules in the arithmetic logic units (ALU). A half adder adds two inputs A and B and produces two outputs as SUM and CARRY, and a conventional CMOS-based half adder consists as many as 18 transistors[14]. As shown in Figure **2a**, the simplest half adder design incorporates an XOR gate for SUM and an AND gate for CARRY. Unlike the linearly separable logic gates shown in Figure **1**, the XOR gate is a linearly inseparable logic function that requires to define two logic thresholds for device with a monotonic input-output response, and thereby hardly possible to be implemented by a single device or simple circuits.

Nevertheless, the XOR gate can be formed by an OR gate and a NAND gate. Following this way, two complementary -*y* and +*y* side laser annealed LSOT devices, denoted respectively as P and Q in Figure **2b**, were connected for the XOR implementation. When programming the initialization currents of P and Q to be $I_{init}^{P}$ = +8 mA and $I_{init}^{Q}$ = -8 mA, OR and NAND gates were obtained, respectively, and the synthetic output of ($V_{Hall}^{P}$ + $V_{Hall}^{Q}$) was shown in Figure **2c**. Binary outputs of around 0 (defined as logic value '0' here) or 40 μV (logic value '1') were found, corresponding to the resulting magnetization states of either P being spin down/up and Q being spin up/down or both P and Q being spin up. Thus, nonlinear separated logic gate of XOR was realized, which can work as SUM for a half adder. Together with another -*y* side laser annealed LSOT device denoted as S, which performed AND function and act as CARRY under a negative initialization current $I_{init}^{S}$ = -8 mA, a spin-orbit half adder was successfully proposed by only three LSOT devices.

# 4. Initialization-free, working current-programmable stateful logics for in-memory computing

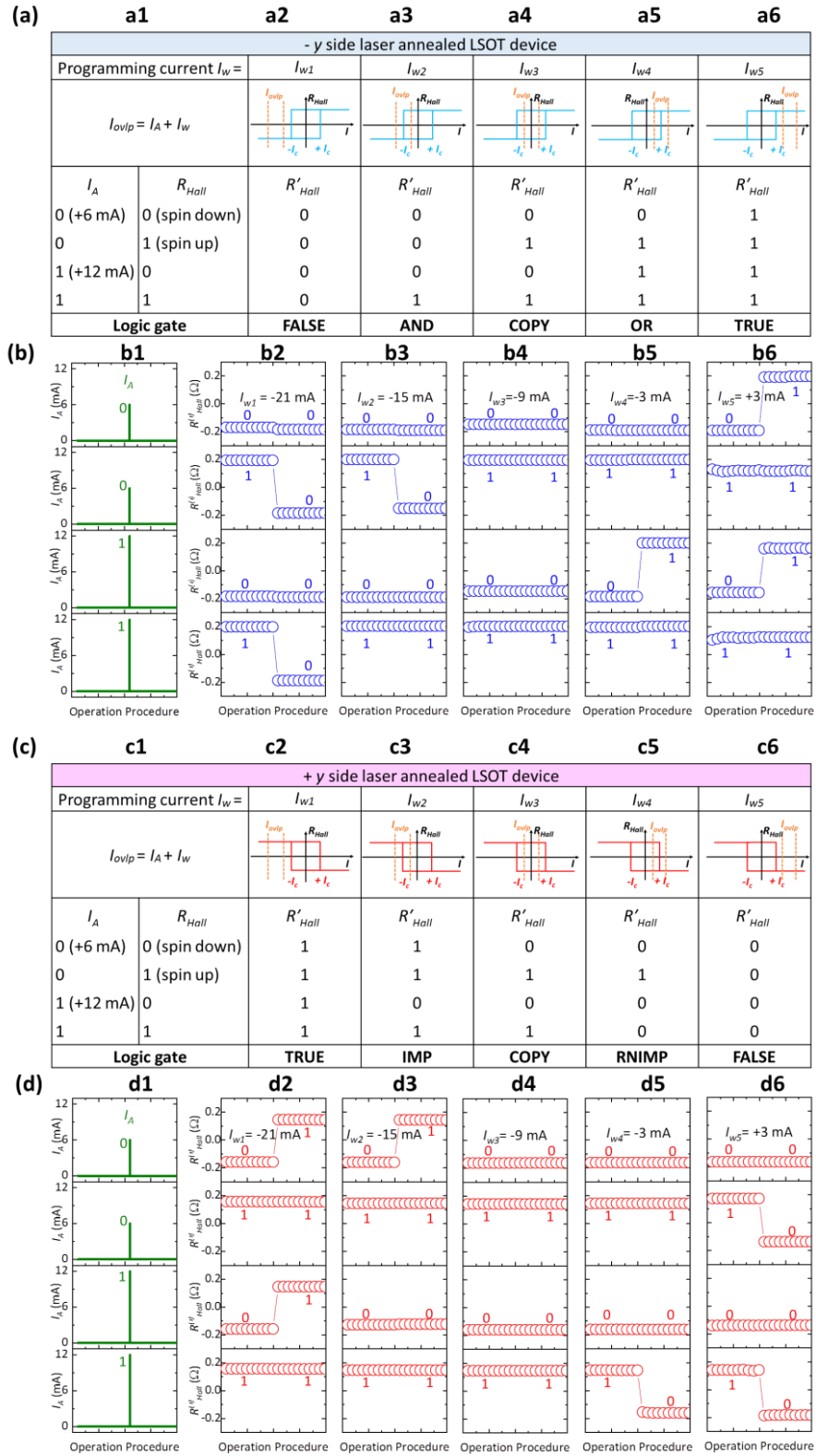

Figure 3 | Design and experimental demonstration of initialization-free, working

**current-programmable stateful _IR-R_ logic operations with the complementary LSOT devices.** A current pulse $I_A$ (here +6/+12 mA stands for current input '0'/'1') and the initial magnetization state (represented by anomalous Hall resistance $R_{Hall}$, and a negative/positive $R_{Hall}$ stands for logic value '0'/'1') were employed as two logic inputs while the resulting magnetization state (represented by $R'_{Hall}$) _in situ_ stored the logic output. In the meantime, a working current pulse $I_w$ was also applied to programming the logic gate types. **a,** Proposal and **b,** respective experimental demonstration for five $I_w$-programmable logic gates of (**a2, b2**) FALSE with $I_{w1}$ = -21 mA, (**a3, b3**) AND with $I_{w2}$ = -15 mA, (**a4, b4**) COPY with $I_{w3}$ = -9 mA, (**a5, b5**) OR with $I_{w4}$ = -3 mA, and (**a6, b6**) TRUE with $I_{w5}$ = +3 mA based on the -_y_ side laser annealed device. **c,** Proposal and **d,** respective experimental demonstration for five corresponding $I_w$-programmable logic gates of (**c2, d2**) TRUE, (**c3, d3**) IMP, (**c4, d4**) COPY, (**c5, d5**) RNIMP, and (**c6, d6**) FALSE based on the +_y_ side laser annealed device. Inserts of (**a**) and (**c**) show schematic drawings of the $R_{Hall}$-_I_ loops with relationship between the overlapped current $I_{ovlp} = I_A + I_w$ (yellow dashed lines) and the critical switching current $\pm I_c$. The _x_-axis of **b** and **d** are operation procedures with same scales and values.

Note that although the output data were stored within the nonvolatile LSOT devices in the form of magnetization states (and thereby can be presented by anomalous Hall resistances or tunneling resistances of MTJs), all electrical input variables ($I_A$ and $I_B$) were used for the above demonstrated logic functions, which were referred to *I-R* logics. However, future in-memory computing, which aims to eliminate the memory wall problem[24] in von-Neumann computing architecture, requires *R-R* logic gates that the processing devices can not only store output data but also perform stateful logic operations by regarding their initial states as input variables at the same time[25-28]. In the following section, we will firstly demonstrate stateful *IR-R* logic gates based on the complementary LSOT devices, where the current pulse $I_A$ and the anomalous Hall resistance $R_{Hall}$ act as two input variables while the resulting $R'_{Hall}$ is stored as the output. Then, by proposing connected LSOT-MTJ circuits, we will show how the demonstrated *IR-R* gates can be converted into equivalent cascading fully tunneling resistive *RR-R* gates for practical in-memory processing.

A key difference between the *I-R* and the *R-R* logic gates is the role of the initial magnetization state, which act as the programming term and the input variable for the *I-R* and the *R-R* gates, respectively. On the one hand, this makes stateful *R-R* logic gates naturally free from initialization operations; on the other hand, other programming methods have to be involved for assembling multi-functional *R-R* gates in a single device, or the device would only perform as one specific gate. As shown in Figure **3**, a working current pulse $I_w$ was simultaneously applied with $I_A$ to program the overlapped $I_{ovlp} = I_A + I_w$. Particularly, five working modes with respective relationships between $I_{ovlp}$ and the critical switching current $\pm I_c$ shown in Figure **3a** and **3c** were derived for $I_A$ = +6 mA (as logic value '0') or + 12 mA (as logic value '1'). Considering the complementary switching senses for the *-y* and the *+y* side laser annealed LSOT devices, more $I_w$-programmable (with $I_w$ = -21, -15, -9, -3, or +3 mA) stateful logic gates, including FALSE, AND, COPY $R_{Hall}$, OR, TRUE, material implication ($I_A$ IMP $R_{Hall}$), and reverse nonimplication ($I_A$ RNIMP $R_{Hall}$) were designed and experimental realized, as shown in Figure **3b** and **3d**.

The working paradigm of above *IR-R* spin logic gates is thought to be applicable for other types of current-driven magnetization switching devices as well. However, advantages of the complementary LSOT devices used here should be underlined due to their capability of significantly enriching the in-memory functionalities and thereby fabricating more straightforward circuits. For example, a -*y* side laser annealed LSOT device can act as an in-memory 3-input majority gate (MAJ) [10] if the $I_w$ is also considered as a logic input. A MAJ returns '1' if and only if the majority (more than half) of its inputs are '1'. Refer to the AND gate ($I_A \wedge R_{Hall}$, abbr. $I_A R_{Hall}$) as shown in Figure **3b3** and the OR gate ($I_A \vee R_{Hall}$) as shown in Figure **3b5**, when $I_w$ = -15/-3 mA is defined as logic input '0'/'1', the logic output can be expressed as $R'_{Hall} = \neg I_w(I_A R_{Hall}) \vee I_w (I_A \vee R_{Hall}) = I_A I_w \vee R_{Hall} I_A \vee R_{Hall} I_w = <R_{Hall}, I_A, I_w>$, where "$<>$" is the logic operator for MAJ. Together with the +*y* side laser annealed LSOT device, which can be programmed to the functionally complete IMP gate[29] as shown in Figure **3d3**, spin-orbit in-memory computing circuit designs with versatile reconfigurable operations are promising.

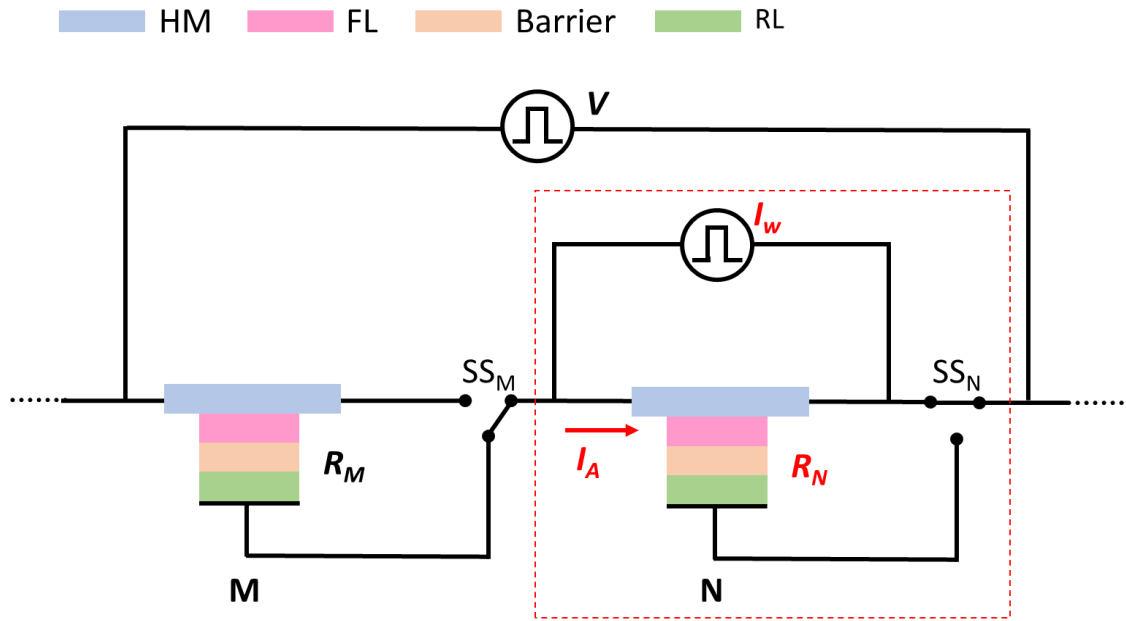

**Figure 4 | Schematic drawing of the cascading stateful *RR-R* logic proposed in connected complementary LSOT-MTJs.** From top to bottom, the LSOT-MTJ consists of a heavy metal electrode (HM), a magnetic free layer (FL), an insulating barrier, and a magnetic reference layer (RL). For every LSOT-MTJ (for example, M), a binary selector switch (SS) is designed to connect either its writing channel HM or its reading channel RL to the HM of next LSOT-MTJ (N). When a constant voltage pulse *V* was applied on M and N, meanwhile $SS_M$ and $SS_N$ were respectively turned to M-reading channel and N-writing channel, the magnetization state of M-FL can be converted into an input current $I_A$ acting on N. Thus, cascading stateful *RR-R* logic that both $R_M$ and $R_N$ work as the input while $R_N$ *in situ* stores the output can be expected.

Actually, *IR-R* gates are equivalent to cascading *RR-R* gates in resistive MTJ devices, where the input current $I_A$ can be converted from tunneling magnetoresistance (TMR) of the preceding device. As shown in Figure **4**, a chain of several complementary LSOT-MTJs was proposed to show cascading *RR-R* logics by connecting each LSOT-MTJ with a binary selector switch (SS). The SS selects which channel of the 3-terminal MTJ, i.e. the in-plane spin-orbit writing channel (heavy metal electrode, HM) or the out-of-plane tunneling reading channel (magnetic reference layer, RL), to be conductively connected with the writing channel of next MTJ. In the case shown in Figure **4**, for instance, $I_w$-programmable *RR-R* logic operations acting on N utilize both magnetization states of the M and N as the logic inputs without changing the magnetization state of M, and thereby such programable stateful in-memory computing can cascade to the rest of LSOT-MTJs of the chain as well.

## 5. Conclusion

In summary, a pair of magnetic field-free complementary LSOT devices with opposite current-driven magnetization switching senses were demonstrated as building blocks for programmable logic and in-memory computing. 4 single-device *I-R* logic gates of AND, NAND, OR, and NOR based on the 2 complementary LSOT devices were firstly obtained by applying two simultaneous current pulses as the logic inputs, before which the device was pre-programmed by an initialization current. A spin-orbit half adder consisting of three LSOT devices was then demonstrated, within which the nonlinear separated logic gate of XOR was realized by a combination of two complementary LSOT devices. After that, initialization-free stateful *IR-R* logic gates of TRUE, FALSE, COPY, AND, OR, IMP, and RNIMP were experimentally demonstrated by regarding the initial magnetization state as one of the logic input, where an additional working current was applied to program the logic modes. Finally, by separating selection of the reading and the writing channels of a MTJ, cascading *RR-R* logic operations in a chain of 3-terminal complementary LSOT-MTJs were proposed. Considering the feasibility of fabricating such complementary LSOT devices by localized laser annealing, the demonstrated pair of building blocks provide an integration-friendly way towards scalable and efficient programmable spin-orbit logics and future in-memory computing.

**Experimental Section**

*Sample preparation:* The Pt(3 nm)/Co(0.5 nm)/Pt(2 nm) layers were deposited onto 0.5 mm-thick Si wafers with a 190 nm-thick thermal $SiO_2$ surface by d.c. magnetron sputtering. The base pressure of the chamber was less than $2 \times 10^{-8}$ Torr, and the pressure of the chamber was 0.8 mTorr under Ar pressure during deposition. The deposition rates of Pt and Co were controlled to be ~0.023 nm $s^{-1}$ and ~0.012 nm $s^{-1}$, respectively. The film was patterned into Hall bar devices using standard photolithography and lift-off processes. The width of the Hall bar is 4 μm. A laser with wavelength of 532 nm and power of 10 mW was used to locally anneal each center region of Hall bar in air atmosphere by sweeping across it along the *x*-direction with a velocity of 0.167 μm/s, leaving a localized laser annealing track on the −*y* or +*y* side. The sweeping was realized through fixing the laser spot position meanwhile controlling the position and movement of the sample by a three-dimensional automated stage and a Thorlabs APT piezo controller, with an in-plane resolution of 5 nm.

*Measurement:* The current-induced magnetization switching and logic operations were carried out using a Keithley 2602B as the current source and Keithley 2182 as the nanovoltmeter. For logic operations, the Keithley 2602B provided an overlapped current $I_{ovlp}$ which is the arithmetic sum of two current signals $I_A$ and $I_B$ (or $I_w$). All measurements were carried out at room temperature without any external magnetic field.

**Acknowledgements**

This work was supported by National Key R&D Program of China (Grant No. 2017YFB0405700), by the NSFC (Grant No. 11474272 and 61774144), by Chinese Academy of Sciences (Grant No. QYZDY-SSW-JSC020, XDPB12, and XDB28000000), and by Beijing Natural Science Foundation Key Program (Grant No. Z190007). We also acknowledge the support from K. C. Wong Education Foundation.

**Competing Interests**

The authors declare no competing financial interests.

**Additional Information**

Correspondence and requests for materials should be addressed to K.W.